\documentclass[preprint,epsfig,aps]{revtex4}
\usepackage{epsfig}
\begin{document}

\title{The di-$\Omega$ in the extended 
quark delocalization, color screening model}

\author{Hourong Pang}
\affiliation{Department of Physics, 
Nanjing University, Nanjing, 210093, P. R. China}
\author{Jialun Ping}
\affiliation{Department of Physics, Nanjing Normal University, Nanjing, 
210097, P.R. China;\\
Center for Theoretical Physics, Nanjing University, Nanjing, 210093, P.R.
China}
\author{Fan Wang}
\affiliation{Center for Theoretical Physics and Department of Physics, 
Nanjing University, Nanjing, 210093, P. R. China}
\author{T. Goldman}
\affiliation{Theoretical Division, Los Alamos National Laboratory, 
Los Alamos, NM 87545, USA}

\begin{abstract} 
The $\Omega\Omega$ (SIJ=-6,0,0) dibaryon state is studied with the
extended quark-delocalization color-screening model including a
$\pi$ meson exchange tail, which reproduces the properties of
the deuteron quantitatively.  We find the mass of the di-$\Omega$ to be
about $45$ MeV lower than the $\Omega-\Omega$ threshold. The effect of
channel coupling due to the tensor force has been calculated and found
to be small in this case. We have also studied the effect of other
pseudoscalar meson exchanges and sensitivity to the short-range cutoff
radius, $r_0$, for the meson exchanges.
\end{abstract}


\maketitle

\section{Introduction}
Quantum ChromoDynamics (QCD) is believed to be the fundamental theory
of the strong interaction. However it is difficult to use to calculate
the low energy properties of complicated quark-gluon systems directly,
such as the NN interaction and the structure of multi-quark-gluon
systems.  Various quark models afford a framework to understand the
physics in these cases.

In the last 30 years, various QCD inspired models have been developed
with both successes and failures in explaining low energy
hadron physics.  To mention a few but certainly not all, there are the
MIT bag model\cite{MITbag}, cloudy bag model\cite{tony}, Friedberg-Lee
nontopological soliton model\cite{flnts}, Skyrme topological soliton
model\cite{sts}, the constituent quark model\cite{cqm}, etc. Different
models use quite different effective degrees of freedom, which might be
indicative of the nature of low-energy QCD. Under different
approximations one can "derive" these models from QCD\cite{cahill}.
Recently there has been a hot debate regarding the proper effective
degrees of freedom of the constituent quark model\cite{isgur}.

We performed a phenomenological study of the baryon interactions with
three constituent quark models, the Glozman-Riska-Brown quark-meson
coupling model\cite{glozman}, the Fujiwara quark-gluon-meson coupling
model -- one version of the Manohar-Georgi chiral quark
model\cite{fujiwara}, and the quark delocalization, color screening
model(QDCSM) developed by ourselves -- a modified version of the de
Rujula-Georgi-Glashow-Isgur quark gluon coupling
model\cite{rggi,wang,eprint}, and found that, in about 2/3 of the
channels formed from the octet and decuplet baryons, these three models
gave qualitatively the same effective baryon-baryon (B-B)
interactions\cite{pang}. These constituent quark models have the same
long range Goldstone boson exchange interaction, (except for the QDCSM,) 
but have different mechanisms for the intermediate and short range
interactions when they are used to study B-B interactions. Present data
on baryon spectroscopy and B-B interactions seem to be insufficient to
distinguish between these models and they all seem to have an
approximate QCD basis\cite{cahill}.  A new generation of hadron
spectroscopy should be helpful in disentangling these hadronic models.

There are good reasons, both theoretically and phenomenologically, to
believe that a two body confinement potential might be a good
approximation for single hadrons. However there is no compelling reason
to believe it is also a good approximation for multihadron systems.
Dibaryons provide a good testing ground for different confinement
mechanisms and baryon interaction models. Being the unique, stable, B=2
system, and its size, binding energy and quadrupole  moment having been
measured very precisely, the deuteron plays a vital role in the
development of NN interaction models.  Since the H-particle, a dibaryon
with strangeness $S=-2$, isospin I=0 and spin J=0, was first predicted
by the MIT bag model in 1977\cite{jaffe}, tremendous efforts have been
made both experimentally and theoretically to search for it. Up to now
there is no experimental evidence for the H.

Other dibaryon candidates have also been discussed in the literature.
An S=0, $J^{P}=0^-$, M$\sim$2.06 GeV, $\Gamma$$\sim$0.5 MeV, I=0 or 2
dibaryon, $d'$, was a hot topic in the 1990's\cite{bilger,prc62}. The
S=0, I=0, $J^{P}=3^+$ dibaryon, $d^*$, has also been a long standing
topic; almost all quark models predict there should be attraction in
this channel but with differing strengths in the different
models\cite{eprint,kf,prc51,gmp,dstar1}. In 1987, we showed that the
S=-3, I=1/2, J=2 dibaryon state might be a narrow resonance in a
relativistic quark model~\cite{goldman}.  In 1990, V.\ B.\ Kopeliovich
{\it et al}.\cite{kopeliovich} predicted that there are strong
interaction stable dibaryons with high strangeness, such as an S=-6,
di-$\Omega$ within the flavor SU(3) Skyrmion model. A general survey of
dibaryon states carried out with both the nonrelativistic QDCSM and the
relativistic version, using an adiabatic calculation, found few high
strangeness states with masses around the lowest
threshold\cite{prc51,gmp}.  Zhang et al.\cite{zhang} predicted that the
$\Omega\Omega$ I=0, J=0 dibaryon is strong interaction stable using a
chiral SU(3) quark model.

In this study, we use the extended QDCSM~\cite{eprint} to calculate the
eigen energy of the di-$\Omega$ state $SIJ=-6,0,0$.  This model allows
the quark system to choose its most favorable configuration through its
own dynamics in a larger Hilbert space than the other models, and takes
into account the possible difference of the confinement interaction
inside a single baryon and between two color singlet baryons. In
particular, this is the unique model that gives an explanation of the
similarity between molecular and nuclear forces.  It also explains why
nuclei are accurately viewed as a collection of nucleons rather than a 
single big bag with 3A quarks.

Here, we recalculate the QDCSM result because the extended QDCSM
reproduces the deuteron properties well quantitatively, unlike the
QDCSM,  and because preliminary NN phase shift calculations using it
fit the NN, N$\Lambda$, N$\Sigma$ scattering data better than the
earlier QDCSM calculations, which fit the data only
qualitatively~\cite{wang,eprint}.  Consequently, we expect this
extended QDCSM calculation will provide a better model estimate of 
the mass of the di-$\Omega$.

In the following section, we give a short description of the extended
QDCSM. The results and a discussion are presented in section III.

\section{Short description of the extended QDCSM}
The details of the QDCSM and its extension can be found in 
Refs.\cite{wang,eprint,prc51} and the resonating-group calculation 
method (RGM) has been presented in Refs.\cite{dstar1,Buchmann}. Here 
we present only the (complete) extended model Hamiltonian, wave 
functions and the necessary equations used in the current calculation.

The Hamiltonian for the 3-quark system is the same as the usual quark 
potential model. For the six-quark system, it is assumed to be
\begin{eqnarray} 
H_6 & = & \sum_{i=1}^6 (m_i+\frac{p_i^2}{2m_i})-T_{CM} +\sum_{i<j=1}^{6} 
    \left( V_{conf}(r_{ij}) + V_G(r_{ij}) +V_{\chi}(r_{ij}) \right) , 
    \nonumber \\ 
V_G(r_{ij}) & = & \alpha_s \frac{\vec{\lambda}_i^c \cdot 
 \vec{\lambda}_j^c }{4} 
 \left[ \frac{1}{r_{ij}}-\frac{\pi \delta (\vec{r})}{m_i m_j} 
 \left( 1+\frac{2}{3} \vec{\sigma}_i \cdot \vec{\sigma}_j \right) 
  + \frac{1}{4m_im_jr^3} \left( 3(\vec{\sigma}_i \cdot 
 \hat{\vec{r}}) (\vec{\sigma}_j \cdot \hat{\vec{r}}) 
- \vec{\sigma}_i \cdot 
  \vec{\sigma}_j \right) \right], \nonumber  \\
V_{\chi}(r_{ij})&=&\{ \sum_{a=1}^{3}V_{\pi}(r_{ij})
{\lambda}_i^{f,a} \cdot {\lambda}_j^{f,a}+\sum_{b=4}^{7}V_{k}(r_{ij})
{\lambda}_i^{f,b} \cdot {\lambda}_j^{f,b}
+V_{\eta}(r_{ij}) \lambda_i^8 \cdot \lambda_j^8 \} 
, \nonumber \\
V_{\gamma}(r_{ij}) & = & \theta (r-r_0) 
\frac{g_8^2}{4\pi}\frac{\mu_{\gamma}^2}
{12m_im_j} 
\frac{1}{r} e^{-\mu_{\gamma} r}  \label{hamiltonian} \\
&&\times \left [\vec{\sigma}_i \cdot \vec{\sigma}_j
+\left (\frac{3(\vec{\sigma}_i \cdot \vec{r})(\vec{\sigma}_j \cdot 
\vec{r})} {r^2} - \vec{\sigma}_i \cdot \vec{\sigma}_j \right ) \left 
(\frac{3}{{(\mu_{\gamma}r)}^2}
+\frac{3}{\mu_{\gamma} r} +1 \right ) \right ] , \hspace*{0.2in} 
\gamma= \pi, K, \eta   \nonumber  \\ 
&&\frac{g_{qq\pi}^2}{4\pi}={(\frac{3}{5})}^2\frac{g_{NN\pi}^2}{4\pi}, 
\nonumber \\
&&\frac{g_8^2}{4\pi}={(\frac{3}{5})}^2\frac{g_{NN\pi}^2}{4\pi}
{(\frac{m_q}{m_N})}^2
\nonumber \\
V_{conf}(r_{ij}) & = & -a_c \vec{\lambda}_i \cdot \vec{\lambda}_j 
\left\{ \begin{array}{ll} 
 r_{ij}^2 & 
 \qquad \mbox{if }i,j\mbox{ occur in the same baryon orbit}, \\ 
 \frac{1 - e^{-\mu r_{ij}^2} }{\mu} & \qquad
 \mbox{if }i,j\mbox{ occur in different baryon orbits}, 
 \end{array} \right. \nonumber \\
\theta (r-r_0) & = & \left\{ 
 \begin{array}{ll}  0 & \qquad r < r_0, \\  1 & \qquad \mbox{otherwise}, 
 \end{array} \right. \nonumber 
\end{eqnarray} 
where $\vec{\sigma}_i$, $\vec{\lambda}_i^c$ and $\vec{\lambda}_i^f$
are, respectively, the spin, color and flavor operators of the $i^{th}$
quark. The spin operators are represented by the three $2 \times 2$
Pauli matrices, and the color and flavor operators are represented by
the eight $3 \times 3$ Gell-Mann matrices. The $\mu_{\gamma}$ are the
masses of the Goldstone bosons ($\gamma=\pi, K, \eta$).  We assume a
single quark-meson coupling constant $g_8^2/4\pi$ for all mesons.  The
confinement potential $V_{conf}(r_{ij})$ has been discussed in
Refs.\cite{dstar1,prc62}.

We obtained the values of b, $\alpha_s$ and $a_c$ by reproducing the
$N-\Delta$ mass difference, the nucleon mass and applying a stability
condition. The mass of the u, d quark is assumed to be 1/3 of the
nucleon mass to meet the requirements of the resonating group method
(RGM) calculation while the strange quark mass $m_s$ is determined by
an overall fit to the baryon octet and decuplet. A flavor symmetric 
overall octet quark-meson coupling constant is assumed for all of the 
octet mesons ($\pi, K, \eta$). 

The color screening parameter $\mu$ is determined by the binding energy
of the deuteron.  The full set of parameters reproduces other
properties of deuteron as well: the size and  the D-S wave mixing
amplitude.  It is worth emphasizing that the extended QDCSM has only
one parameter, the color screening constant $\mu$, that needs to be
adjusted to obtain the correct binding energy of the deuteron, and that
this is sufficient to reproduce well all of the properties of the
deuteron. The tensor force of pion exchange plays a vital role in
reproducing the deuteron properties quantitatively, but the results are
not extremely sensitive to the meson exchange cutoff parameter,
$r_{0}$.

After introducing Gaussian functions with different reference
centers $S_i$, i=1...n, which play the role of the generating
coordinates in this formalism, and including the wave function for the
center-of-mass motion\footnote{For details, see
Refs.\cite{dstar1,Buchmann}.}, the ansatz for the two-cluster wave
function used in the RGM can be written as
\begin{eqnarray}
\Psi_{6q} & = & {\cal A}  \sum_k \sum_{i=1}^{n} \sum_{L_k=0,2} C_{k,i, L_k} 
  \int \frac{d\Omega_{S_i}}{\sqrt{4\pi}}
  \prod_{\alpha=1}^{3} \psi_{\alpha} (\vec{S}_i , \epsilon) 
  \prod_{\beta=4}^{6} \psi_{\beta} (-\vec{S}_i , \epsilon)  \nonumber \\
  & &  [[\eta_{I_{1k}S_{1k}}(B_{1k})\eta_{I_{2k}S_{2k}}(B_{2k})]^{IS_k} 
Y^{L_k}(\hat{\vec
{S_i}})]^J
   [\chi_c(B_1)\chi_c(B_2)]^{[\sigma]}
	\label{multi}  ,
\end{eqnarray}
\noindent where k is the channel index. For the di-$\Omega$, we have
$k=1,2$, corresponding to the channels $\Omega\Omega$ S=0, L=0 and S=2,
L=2. We consider only these channels as they are the only ones coupled
by the tensor interaction.

The delocalized single-particle wave functions used in the QDCSM are
\begin{eqnarray}
\psi_{\alpha}(\vec{S}_i ,\epsilon) & = & \left( \phi_{\alpha}(\vec{S}_i) 
+ \epsilon \phi_{\alpha}(-\vec{S}_i)\right) /N(\epsilon), \nonumber \\
\psi_{\beta}(-\vec{S}_i ,\epsilon) & = & \left(\phi_{\beta}(-\vec{S}_i) 
+ \epsilon \phi_{\beta}(\vec{S}_i)\right) /N(\epsilon), \nonumber \\
N(\epsilon) & = & \sqrt{1+\epsilon^2+2\epsilon e^{-S_i^2/4b^2}}. \label{1q} \\
\phi_{\alpha}(\vec{S}_i) & = & \left( \frac{1}{\pi b^2} \right)^{3/4}
   e^{-\frac{1}{2b^2} (\vec{r}_{\alpha} - \vec{S}_i/2)^2} \nonumber \\
\phi_{\beta}(-\vec{S}_i) & = & \left( \frac{1}{\pi b^2} \right)^{3/4}
   e^{-\frac{1}{2b^2} (\vec{r}_{\beta} + \vec{S}_i/2)^2}. \nonumber
\end{eqnarray}
The delocalization parameter, $\epsilon$, is determined by the dynamics
of the quark system rather than being one of the fixed (adjusted)
parameters.

From the variational principle, after variation with respect to the
relative motion wavefunction $\chi(R)$ , one obtains the RGM equation
\begin{equation}
\int H(\vec R, \vec{R'}) \chi (\vec{R'}) d\vec{R'} =
E \int N(\vec R,\vec{R'}) \chi (\vec{R'}) d\vec{R'} , \label{RGM}
\end{equation}

With the above ansatz, the RGM equation, Eq.(\ref{RGM}),
becomes an algebraic eigenvalue equation,
\begin{equation}
\sum_{j,k,L_k} C_{j,k, L_k} H_{i,j}^{k',L'_{k'},k,L_k} 
  = E \sum_{j} C_{j,k, L_k} N_{i,j}^{k',L'_{k'}}
   \label{GCM}
\end{equation}
where $N_{i,j}^{k',L'_{k'}}, H_{i,j}^{k,L_k,k',L'_{k'}}$ are the
(Eq.(\ref{multi})) wave function overlaps and Hamiltonian matrix
elements, respectively. By solving the generalized eigen value problem,
we obtain the energies of the 6-quark system and their corresponding
wave functions.

\section{Results and discussion}
Our model parameters, which have been fixed by matching baryon and
deuteron properties and using only $\pi$ meson exchange, are given in
Table I.  For comparison, the results of our earlier
calculation~\cite{prc51} without $\pi$ meson exchange are also included
in Table I.  In order to study the dependence of the result on the
short range cutoff radius, we choose three typical values: 0.6, 0.8 and
1.0 fm, for the short-range cutoff of the $\pi$ meson exchange
potential. This cutoff is necessary in our model approach because all
short and intermediate range effects are already represented by the
quark delocalization and color screening: The short range repulsion of
the NN interaction is provided by a combination of the color magnetic
interaction due to gluon exchange and the Pauli principle enforced by
the quark structure of the nucleon; the intermediate range attraction
conventionally modeled by heavy meson and multimeson exchange is also
provided by quark delocalization and color screening.  If the contact
term of single $\pi$ meson exchange, $\delta(\vec{r})$, or its smeared
version, exp$(-\Lambda r)/ (\Lambda r)$, were used, double
counting\cite{eprint} would occur.  In all cases, we found that the
contribution of $\pi$ meson exchange to the dibaryon mass is not large
(about 10 MeV or less), primarily due to the effect of the short-range
cutoff; correspondingly, the deviations from the original QDCSM
parameters are also quite small.

\begin{center}
Table I. Model parameters and results calculated for di-$\Omega$. 
\begin{tabular}{c|c|c|c|c|c} \hline
 &  \multicolumn{3}{|c|} {including $\pi$ meson exchange} & 
\multicolumn{2}{|c}{excluding meson exchange}  \\ \hline
 & $r_0=0.6$ fm & $r_0=0.8$ fm & $r_0=1.0$ fm & dynamical cal.&
   adiabatic cal.  \\ \hline
 $m_d,m_s$ (MeV) & 313 , 634 & 313 , 634 & 313 , 634 & 313, 634& 313 , 634 \\
 $b$ (fm)& 0.6010 & 0.6015 & 0.6021 & 0.6034 &  0.6034\\ 
 $a_c$ (MeV fm$^{-2}$) & 25.40 & 25.14 & 25.02 & 25.13 & 25.13 \\
 $\alpha_s$  & 1.573 & 1.5585 & 1.550 & 1.543 & 1.543 \\ \hline
 $\mu$ (fm$^{-2}$) & 0.75 & 0.85 & 0.95 & 1.0 & 1.0 \\ \hline
Mass$_{sc}$ (MeV) &  3290.3  &  3298.2  &  3303.5  &  3312 & 3350 \\  \hline
Mass$_{cc}$ (MeV) & 3290.2 & 3297.9 & 3303.2 & - & - \\ \hline
\end{tabular}
\end{center}
$sc$ refers to a $k=1$ single channel calculation, $cc$ refers to the 
coupled channel calculation.

Comparing the last two columns in Table I, where the $\pi$ meson
exchange tails have not been included, we see that the dynamic
calculation modifies the results of the previous adiabatic calculation
significantly. The mass of the $\Omega\Omega$, which is close to the
threshold ($3344.9$ MeV) in the adiabatic approximation, is reduced
below the threshold by about $33$ MeV.  Several effects contribute to
this difference, but the most important one is that the resonating
group method calculates the relative motion between the two quark
clusters rigorously, greatly improving over the rough estimate of the
kinetic energy by the zero-point oscillation energy used in the
adiabatic approximation.

We note that one should not expect all baryon-baryon (B-B) states to be
modified significantly by the dynamical calculation. If the effective
B-B interaction has a narrow minimum with respect to the separation
variable, then the zero-point oscillation energy provides a good
approximation to the kinetic energy of the relative motion of the two
clusters. For example, the $d^*$ mass does not change very much between
the adiabatic and dynamic calculations. However, if the effective B-B
interaction is flat, then the zero-point oscillation energy may be not
a good approximation for the kinetic energy of the relative motion;
the di-$\Omega$ is such a case\cite{pang}.

The first three columns show the dynamical mass of the di-$\Omega$ with
(only) $\pi$ meson exchange included but with different cutoff parameters. 
Comparing these three results with the fourth column, one can see that
the di-$\Omega$ mass is reduced by between 9 and 22 MeV. This is due to
the fact that the parameters of the model are modified by inclusion of
the pion even though the pion is not exchanged between s quarks.  The
binding energies vary from 55~MeV to 40~MeV  with different cutoffs.
The larger the cut off, the smaller is the effect on the di-$\Omega$
mass.

The last two rows show that the effect of channel coupling caused by
the meson tensor force is quite small here.  The S and D wave coupled
channel masses (cc) are almost the same as those of the single channel
approximation (sc). In the deuteron, the effect on the mass of the
state, although almost an order of magnitude larger and more significant 
there, was also found to be small. By contrast, as shown below, the
D-S wave mixing in both cases is comparable.


The quark spin-orbit interaction has been neglected in the Hamiltonian
(1).  We would expect it to have a similarly significant effect on the
D-wave channel. However, its effect on the di-$\Omega$ mass is expected
to be small because the D-wave channel coupling itself only affects the
di-$\Omega$ mass slightly.

Fig.1 shows the relative motion wave functions for two different cutoff
parameters. They are not sensitive to the short-range cutoff parameter,
$r_0$.  Fig.2 shows the effective $\Omega-\Omega$ potential
corresponding to the cutoff $r_0=0.8fm$. The other cutoffs give almost
indistinguishable effective potentials.

As has been mentioned above, the effects of short and intermediate
range interaction arising from $\pi$ meson exchange are represented by
quark delocalization and color screening, so the only long range effect
in our model comes from $\pi$ exchange.  The effects of heavy
pseudoscalar meson ($K$, $\eta$) exchange on the mass of the
di-$\Omega$ have been checked with a representative cutoff of $r_0=0.8
fm.$ The results are presented in Table II.

\pagebreak

\begin{center}
Table II. Mass of di-$\Omega$ for $r_0=0.8fm$ 
and full octet pseudoscalar meson exchange. 
\begin{tabular}{c|c|c} \hline
&only $\pi$-exchange & $\pi,K,\eta$ -exchange \\ \hline
$m_u,m_s$&313,634&313,634 \\ 
b(fm)&0.6015&0.6022 \\ 
$a_c$ (MeV {fm}$^{-2}$)&25.14&25.03 \\ 
$\alpha_s$ (fm$^{-2})$&1.5585&1.555 \\
$\mu$ &0.85&0.9 \\ \hline
$Mass_{Deuteron}$ (MeV)&1876&1875.8 \\
$\sqrt(r^2)$&2.0&1.92 \\
$P_d$&4.53\%&4.92\% \\ \hline
$Mass_{Di-\Omega}$ (MeV) & 3298.2&3300.0 \\ \hline
\end{tabular}
\end{center}

\noindent Repeating the same process mentioned above to fix the
parameters, we found that adding $K$ and $\eta$-exchange affects the
parameters (b, $a_c$, $\alpha_s$) only slightly; the properties of the
deuteron can be reproduced just as well by a very small readjustment
the color screening parameter, $\mu$. Comparing the results in Table I
and II, we find the mass of the di-$\Omega$ is almost unaffected by the
addition of $K$ and $\eta$ exchange. This confirms our expectation that
heavier (shorter ranged) meson exchanges are already represented by
quark effects in our model, so that explicit representation of such
exchanges beyond a cut-off scale are not important in our approach.

In summary, within the framework of the extended QDCSM, using a
resonating group method, we estimate the mass of the di-$\Omega$ to be
about $45$ MeV lower than the $\Omega-\Omega$ threshold.  This mass is
almost unaffected by channel coupling, the value of the short range
cutoff length, $r_0$, and heavier pseudoscalar meson exchanges.  The
inclusion of single pion or heavier meson exchange has only a minor
effect on the di-$\Omega$ mass.  These results are consistent with the
calculations of Kopeliovich and of Zhang et
al.\cite{kopeliovich,zhang}.

We should, however, take note that the precision of any prediction of
the di-$\Omega$ mass is limited by that of the model itself. For
example, in our model approach, the mass of the $\Omega$ itself is
found to be 1651 MeV, or 21 Mev less than the measured value. It could
well be argued that this theoretical mass of the $\Omega$ should be
used to calculate the threshold. We would then conclude that the
di-$\Omega$ mass is no more than $5$ MeV below threshold. From the 
model deviations from experimental values for single baryon masses, 
we estimate a total systematic uncertainty of order 30 MeV as a 
reasonable value to assign to our model. 

Moreover, it is difficult to estimate how much of the model uncertainty
is carried over to the dibaryon calculation for such a high strangeness
quark system, because the model parameters are mainly determined by the
experimental data of the nonstrange sector.  Therefore, when designing
a detector system to hunt for the di-$\Omega$ state, it is
theoretically advisable to take into account the possibility that the
di-$\Omega$ may be either stable or unstable with respect to the strong
interaction.

This work is partly supported by the HIRFL-CSR theoretical center, the
NSF and SED of China, and partly by the US Department of Energy under
contract W-7405-ENG-36.

\vspace*{0.5in}
\noindent
{\large{Figure Captions}}

\noindent
Fig. 1. Cluster relative motion wave function of di-$\Omega$.

\noindent
Fig. 2. S-wave $\Omega-\Omega$ effective potential.

\pagebreak

\begin{figure}[ht]
\epsfig{file=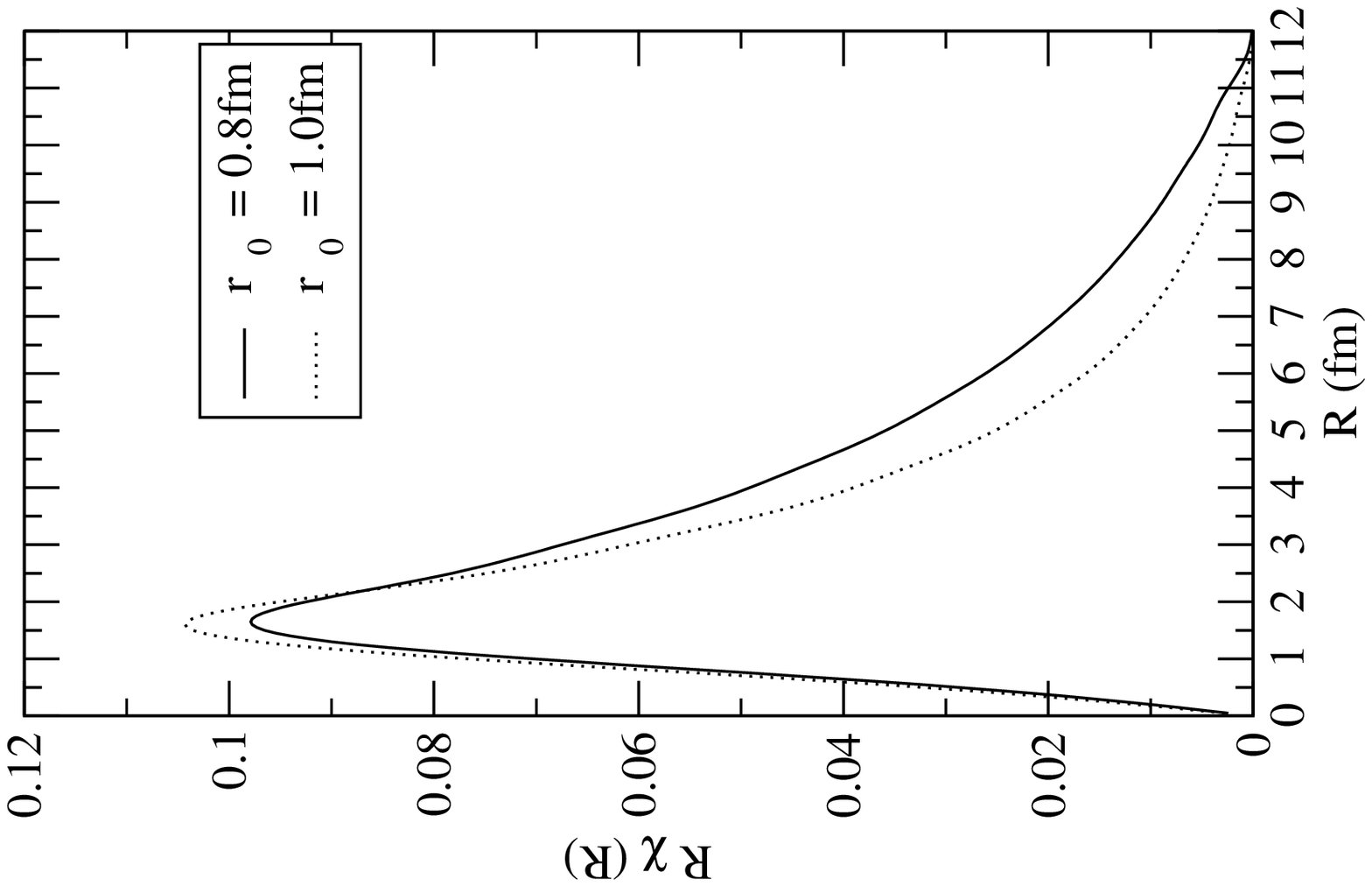,angle=-90}
\caption{Cluster relative motion wave function of di-$\Omega$.
\label{fig1}}
\end{figure}

\pagebreak

\begin{figure}[ht]
\epsfig{file=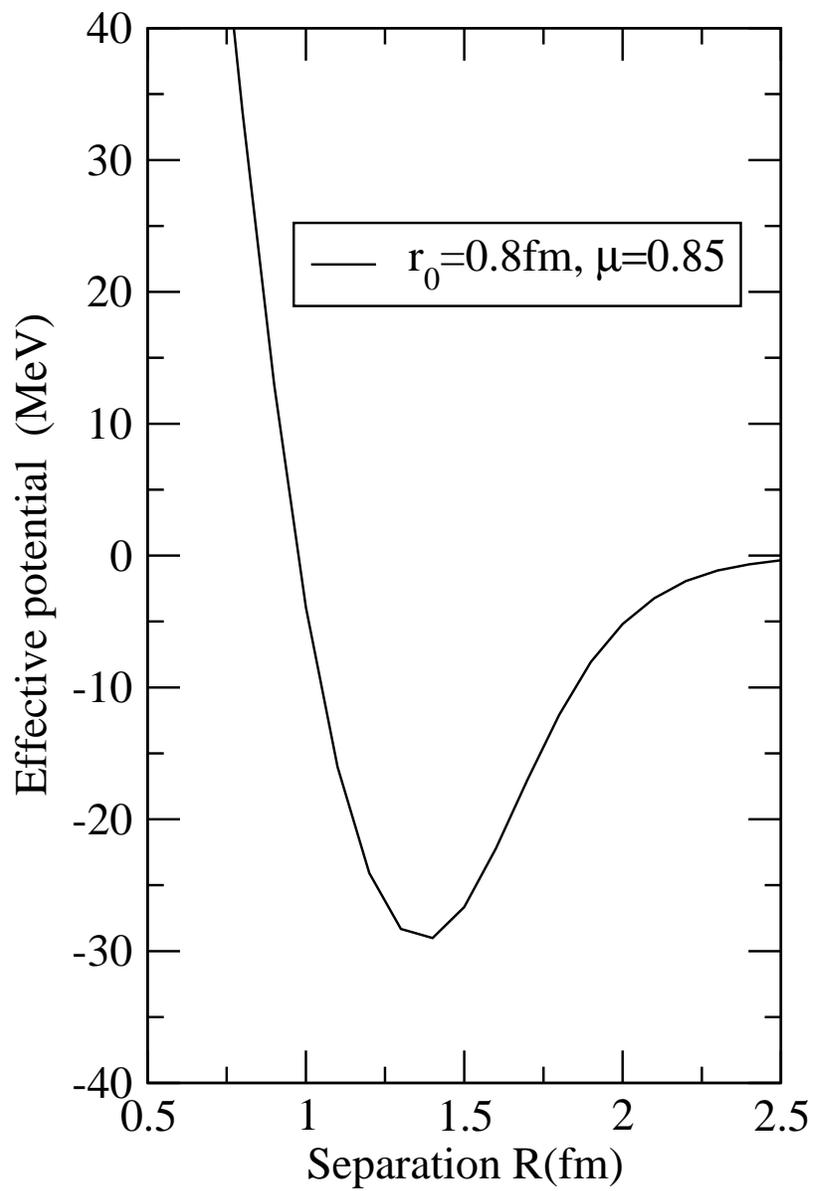,angle=-90}
\caption{S-wave $\Omega-\Omega$ effective potential.\label{fig2}}
\end{figure}

\end{document}